\documentclass[aps,prb,reprint,groupedaddress,amsmath,amssymb,eqsecnum,floatfix]{revtex4-2}

\usepackage[final]{graphicx}
\usepackage{color}
\usepackage[english]{babel}
\usepackage[largesc]{newtxtext}
\usepackage[varg]{newtxmath}
\usepackage{hyperref}

\begin{document}

\title{Out-of-time ordered correlation functions for the localized $f$ electrons in the Falicov-Kimball model}


\author{A. M. Shvaika}
\affiliation{Yukhnovskii Institute for Condensed Matter Physics of the National Academy of Sciences of Ukraine, Lviv 79011, Ukraine}
\author{J. K. Freericks}
\affiliation{Department of Physics, Georgetown University, Washington, DC 20057, USA}

\date{\today}

\begin{abstract}
We provide an exact evaluation of the out-of-time correlation (OTOC) functions for the localized $f$-particle states in the Falicov-Kimball model within  dynamical mean-field theory. Different regimes of quantum chaos and quantum scrambling are distinguished by the winding numbers of the block Toeplitz matrices used in the calculation. The similarities of these fermionic OTOCs and their logarithmic derivatives for time evolution with the OTOCs for quantum spin models with disorder are also discussed.
\end{abstract}


\maketitle


\section{Introduction}

Long-time coherent quantum many-body dynamics is important for condensed matter physics, quantum information science, optical physics, and high-energy physics~\cite{xu_2024:010201,garcia-mata_2023:55237}. First, it is connected with the spreading of quantum information, its detection, and decoding. These dynamics can be examined in a system of qubits whose  Hamiltonian governs unitary dynamics where the initial state of one qubit becomes nonlocal and is distributed across a system by the unitary dynamics---the so-called quantum information scrambling phenomenon. The rate with which information becomes nonlocal is usually quantified by the out-of-time-ordered correlators~\cite{hayden_2007:120,roberts_2015:131603}
\begin{equation}
	F(r,t) = \left\langle [W_0(t),V_r(0)]^{\dagger}[W_0(t),V_r(0)]\right\rangle,
\end{equation}
where $W_0$ and $V_r$ are local operators at positions 0 and~$r$.

Out-of-time-ordered correlators (OTOC) were proposed by Larkin and Ovchinnikov~\cite{larkin_1968:2262,*larkin_1969:1200} as a measure of importance of quantum motion with respect to the quasiclassical approximation in superconductivity. Later, OTOCs were used to describe the evolution of the overlap of perturbed and unperturbed states as indicators of the quantum motion stability in chaotic and regular systems~\cite{peres_1984:1610}.

For chaotic systems, the initially local operator $W_0(t)$ expands, under Heisenberg dynamics, with a speed called the butterfly velocity and overlaps later with the operator $V_r$ resulting from the growth of the OTOC function inside the emergent light-cone in space-time. The growth rate of the OTOC is influenced by the interplay between quantum fluctuations and local scrambling, and depends on the size of the system and on its temperature. For thermal quantum systems with a large number of degrees of freedom, the initial classical exponential growth rate is characterized by the Lyapunov exponent bounded by the temperature $\lambda_L\leq 2\pi k_B T/\hbar$~---this bound is a bound on the growth rate of chaos \cite{maldacena_2016:106,xu_2019:031048}. However, for quantum dissipative systems, e.g. systems in contact with a thermal environment, we find that instead of the short time exponential growth, they exhibit a long time decay with a decay rate that mirrors the classical Lyapunov exponent~\cite{bergamasco_2023:024208,bergamasco_2025:034201}.

In order to estimate the growth rate of an OTOC, in particular, for exponential growth, it is convenient to consider the logarithmic derivative of the OTOC~\cite{lin_2018:144304,riddell_2019:054205,xu_2020:199,buividovich_2022:046001}
\begin{equation}
	L(r,t) = \frac{\mathrm{d} \ln F(r,t)}{\mathrm{d}t} = \frac{1}{F(r,t)} \frac{\mathrm{d} F(r,t)}{\mathrm{d}t},
\end{equation}
which is equal to the Lyapunov exponent for ``fast scramblers'' obeying the  chaos bound, and can provide detailed behavior of the OTOC at its wavefront, even when it is not exponentially growing. Relaxation dynamics in many-body systems can be characterized by different time scales $\tau$, each with their own temperature and wave-vector dependencies, which affect the scrambling, and result in an anomalous behavior for the Lyapunov exponents~\cite{kim_2020:085134}.

For quantum information scrambling, one is interested in the OTOCs of the local two-state operators---that is, for the qubits. As a rule, spin models with a direct interaction between the spins are used to describe their unitary dynamics, see e.g.~\cite{xu_2024:010201} and references therein. On the other hand, in real systems, the qubits are strongly localized and their dynamics, in many cases, are governed by the indirect interaction through other degrees of freedom, which often become retarded and long range.

One of the simplest models with two-level states interacting through (nonlocal) band electrons is the Falicov-Kimball model (FKM)~\cite{falicov_1969:997}, which considers localized $f$-states which locally interact with itinerant $d$-electrons. It was shown by Brandt and Schmidt~\cite{brandt_1986:45} and Kennedy and Lieb~\cite{kennedy_1986:320} that the ground state phase diagram of the FKM can be obtained from the Ising-like (lattice gas) model with an effective interaction between classical spins (occupation of $f$ states) after integrating out the fast $d$-electrons, and the cases of 1D and 2D lattices have been examined in detail~\cite{kennedy_1994:901,gruber_1997:57,datta_1999:545,wojtkiewicz_2001:233103}.

In addition, the FKM possesses exact solutions for  infinite dimensional lattices~\cite{brandt_1989:365,freericks_2003:1333} within the dynamical mean-field theory (DMFT)~\cite{metzner_1989:324,metzner_1991:8549,georges_1996:13}. Despite the $f$-electron occupation number being a conserved quantity (an integral of motion), the $f$-electron spectral function is a complicated object which displays nontrivial features~\cite{brandt_1992:297,freericks_2005:115111,shvaika_2008:425}. The local $d$-electron density-density OTOC for the FKM was calculated and its enhancement near the metal-insulator transition point was obtained~\cite{tsuji_2017:011601}. But, since the $f$-electron density-density OTOC for the FKM is exactly equal to zero, one must look at OTOCs constructed from  $f$-electron creation and annihilation operators, which describe transitions between local $f$-states, similar to the spin-flip operators for spin systems.

In this article, we investigate OTOC functions constructed from $f$-electron creation and annihilation operators (state-flip operators for the two-level system) for the FKM on an infinite dimension hypercubic lattice. First, we introduce the model and recall the derivation of the $f$-electron DOS. Next, we consider the lattice gas model and obtain analytic results for the single-particle DOS and OTOCs for any lattice, and then examine the infinite-dimensional limit. Finally, we calculate OTOCs for $f$-electrons for the FKM and provide numerical results.

\section{Model Hamiltonian and single-particle propagators}

The Falicov-Kimball model (FKM)~\cite{falicov_1969:997} is the simplest model for strongly correlated electron systems and describes a system of the itinerant $d$-electrons which interact with localized particles. Its Hamiltonian
\begin{equation}\label{eq:H_FK}
	H = - \sum_{ij} t_{ij} d_i^{\dagger} d_j + \sum_i H_i 
\end{equation}
contains a hopping contribution from the itinerant electrons (kinetic energy) and a local part
\begin{equation}\label{H_FKM}
H_i = U n_{id} w_{i1} - \mu n_{id} + (E_f - \mu) w_{i1} 
\end{equation}
which includes the interaction of strength $U$ between the itinerant and localized particles, Here, $n_{id} = d_i^{\dagger} d_i$ is the local density of $d$ electrons and $w_{i1}$ is the occupation of site $i$ by the localized particles, which can take two values, 0 or 1, and describes a two-level state subsystem without specifying its internal dynamics. As a rule, it is associated with the localized $f$-electron state, i.e.,
it is the expectation value of the localized electron density operator
\begin{equation}
w_{i1} = n_{if} = f_i^{\dagger} f_i
\end{equation}
and its internal dynamics is determined by the $f$-electron creation $f_i^{\dagger}$ and annihilation $f_i$ operators. These local $f$-electron densities are conserved in the model, as they commute with the Hamiltonian.

The FKM possesses an exact solution in the limit of infinite dimensions $D\to\infty$, after scaling the nearest-neighbor hopping by $t_{ij} = t^{*}/2\sqrt{D}$, and can be solved exactly with DMFT~\cite{georges_1996:13,freericks_2003:1333}. Below, we shall put $t^{*}=1$ as the energy scale.

In DMFT, the self-energy is local~\cite{metzner_1989:324, metzner_1991:8549}, and the lattice model is mapped onto an effective impurity problem with the same self-energy and some auxiliary self-consistent $\lambda$-field~\cite{brandt_1989:365}. Both the self-energy $\Sigma(\omega)$ and $\lambda$-field $\lambda(\omega)$ are found from solutions of the self-consistent set of equations
\begin{align}
	G_d(\omega) &= \frac{1}{\omega+\mu-\Sigma(\omega)-\lambda(\omega)} 
	\nonumber \\
	&= \int_{-\infty}^{+\infty} \mathrm{d} \varepsilon \frac{\rho(\varepsilon)}{\omega+\mu-\Sigma(\omega)-\varepsilon},
	\label{eq:DMFT}
\end{align}
where $G_d(\omega)$ is the $d$-electron Green's function for the impurity problem and for the FKM it becomes
\begin{align}
	G_d(\omega) &= \frac{w_0}{\omega+\mu-\lambda(\omega)} + \frac{w_1}{\omega+\mu-U-\lambda(\omega)}
	\nonumber \\
	&= w_0 \mathcal{G}_0(\omega) + w_1 \mathcal{G}_1(\omega).
	\label{eq:Gd_imp}
\end{align}
Here, $w_1 = n_f = \langle f^{\dagger}f\rangle$ and $w_0 = 1 - n_f = \langle ff^{\dagger}\rangle$ are average occupations of the localized $f$-states on the lattice and $\rho(\omega)$ is the unperturbed density of states, which for the infinite-dimensional hypercubic lattice is a Gaussian
\begin{equation}\label{eq:rho_gauss}
	\rho(\omega) = \frac{1}{t^{*}\sqrt{\pi}} \mathrm{e}^{-(\omega/t^{*})^2}.
\end{equation}
Finally, for the $d$-electron propagator we find
\begin{equation}\label{eq:dGF}
	G_{d,\mathbf{k}}(\omega) = \frac{1}{\omega+\mu-\Sigma(\omega)-\varepsilon_{\mathbf{k}}},
\end{equation}
where $\varepsilon_{\mathbf{k}} = -\frac{t^*}{\sqrt{D}}\sum_{\alpha} \cos k_{\alpha} a$ is the unperturbed band energy.

Finding the solution for the $f$-electron propagator is more challenging~\cite{brandt_1992:297,zlatic_2001:1443,freericks_2003:1333}. We follow the derivation  in~\cite{shvaika_2008:425}.
The time evolution of the $f$-operators is given by
\begin{align}
f_i(t) &= \mathcal{U}_i(t,0) f_i(0),
\label{eq:f_t}
\\
f_i^{\dagger}(t) &= f_i^{\dagger}(0) \mathcal{U}_i^{\dagger}(t,0)
= \tilde{\mathcal{U}}_i(t,0) f_i^{\dagger}(0),
\label{eq:fd_t}
\end{align}
where the evolution operators are
\begin{align}
\mathcal{U}_i(t,0) &= \mathcal{T}\!\exp\left\{-\mathrm{i}\int_{0}^{t}\mathrm{d}t'\left[U n_{id}(t')+E_f-\mu\right]\right\},
\label{eq:Udef}
\\
\tilde{\mathcal{U}}_i(t,0) &= \mathcal{T}\!\exp\left\{\mathrm{i}\int_{0}^{t}\mathrm{d}t'\left[U n_{id}(t') + E_f - \mu\right]\right\}.\label{eq:tUdef}
\end{align}
Based on this, the contour-ordered single-particle propagator is defined to be
\begin{equation}
G_{ijf}^c(t,t')=-\mathrm{i} \left\langle\mathcal{T}_c f_i(t) f_j^\dag(t') \right\rangle =\delta_{ij} G_{f}^c(t,t'),
\label{eq:Gfc_def}
\end{equation}
where the time ordering is performed on the Keldysh contour and the Coulomb interaction $U$ acts only on the upper (forward) branch of contour. It is a local quantity because there is no hopping for the $f$ electrons. In particular, the greater $G_f^{>}(t)$ and lesser $G_f^{<}(t)$ Green's functions can be expressed in terms of functional determinants of Toeplitz matrices calculated on the time interval $[0,t]$ with $t>0$ (for details, see \cite{shvaika_2008:425})
\begin{align}
G_f^{>}(t)&= -\mathrm{i} \left\langle \mathcal{U}_i(t,0) f(0) f^\dag(0) \right\rangle
\nonumber \\
&=-\mathrm{i} w_0 \mathrm{e}^{-\mathrm{i}(E_f - \mu)t} 
\nonumber \\
&\times\det\nolimits_{[0,t]}
\left\| \delta(t_1-t_2) - \mathcal{G}_{0}(t_1-t_2) U \right\|
\label{eq:Gfgr_def}
\end{align}
and [$G_f^{<}(t)=-\{G_f^{<}(-t)\}^*$]
\begin{align}
G_f^{<}(-t)&= \mathrm{i} w_1 \mathrm{e}^{\mathrm{i}(E_f - \mu) t}
\nonumber \\
&\times \det\nolimits_{[0, t]}
\left\| \delta(t_1-t_2) + \mathcal{G}_{1}(t_1-t_2) U \right\|.
\end{align}
Here 
\begin{align}
\mathcal{G}_0(t'-t'') &= - \frac{\mathrm{i}}{\pi} \int_{-\infty}^{+\infty}\mathrm{d}\omega\, \mathrm{e}^{-\mathrm{i}\omega(t'-t'')} \left[f(\omega)-\Theta(t'-t'')\right]
\nonumber \\ 
&\times\operatorname{Im} \frac{1}{\omega + \mu -\lambda(\omega)}
\label{eq:iG0}
\end{align}
and
\begin{align}
\mathcal{G}_1(t'-t'') &= - \frac{\mathrm{i}}{\pi} \int_{-\infty}^{+\infty}\mathrm{d}\omega\, \mathrm{e}^{-\mathrm{i}\omega(t'-t'')} \left[f(\omega)-\Theta(t'-t'')\right]
\nonumber \\ 
&\times \operatorname{Im} \frac{1}{\omega + \mu - U -\lambda(\omega)}
\label{eq:iG1}
\end{align}
are the time ordered (causal) impurity Green's functions for the empty and filled $f$ states, respectively, using the boundary condition for the step function, given by $\Theta(0)=0$~\cite{pakhira_2019:125137}. 
The retarded Green's functions is defined in the usual way via
\begin{align}
G_f^r(t)&=\Theta(t)\left\{G_f^{>}(t)-G_f^{<}(t) \right\}
\nonumber \\
&=-\mathrm{i} \Theta(t) \mathrm{e}^{-\mathrm{i}(E_f - \mu)t} \bigl\{
w_0 \det\nolimits_{[0,t]} \left\| \mathbf{I} - \mathcal{G}_{0} U \right\|
\nonumber \\
&+ w_1 \left(\det\nolimits_{[0,t]} \left\| \mathbf{I} + \mathcal{G}_{1} U \right\| \right)^*
\bigr\}
\nonumber \\
&= G_0^f(t)+ G_1^f(t)
\label{eq:Grt}
\end{align}
with a Fourier transform
\begin{align}
G_f^r(\omega)&=-\mathrm{i} \int_0^{+\infty}\mathrm{d} t\, \mathrm{e}^{\mathrm{i}(\omega + \mu -  E_f)t}
\bigl\{
w_0 \det\nolimits_{[0,t]} \left\| \mathbf{I} - \mathcal{G}_{0} U \right\|
\nonumber \\
&+ w_1 \left(\det\nolimits_{[0,t]} \left\| \mathbf{I} 
+ \mathcal{G}_{1} U \right\| \right)^*
\bigr\},
\label{Gr_w_total}
\end{align}
which defines the single-particle DOS as
\begin{equation}
A_f(\omega) = -\frac{1}{\pi} \operatorname{Im} G_f^r(\omega).
\end{equation}

The long time asymptotic of the Toeplitz determinants in \eqref{eq:Grt} can be obtained by using  Szeg\H{o}'s theorem and the Wiener-Hopf approach, see \cite{shvaika_2008:425} and references therein. It is determined by the winding number (index) of the characteristic functions $C_{0}^f(\omega)$ and $C_{1}^f(\omega)$, which are given by
\begin{align}
&C_{0,1}^f(\omega) = 1 \mp U \mathcal{G}_{0,1}^t(\omega)
\nonumber \\
&= 1 \mp U \operatorname{Re}\mathcal{G}_{0,1}^r(\omega) \mp \mathrm{i} \, U\tanh\frac{\beta\omega}{2}\operatorname{Im}\mathcal{G}_{0,1}^r(\omega),
\label{eq:Cfw}
\end{align}
respectively.
Here, $\mathcal{G}_{0,1}^t(\omega)$ is the Fourier transform of the time-ordered impurity Green's function in \eqref{eq:iG0} and \eqref{eq:iG1}. We choose both time indices $t'$ and $t''$ to be on the top branch of the Keldysh contour, and $\mathcal{G}_{\alpha}^r(\omega)$ is the retarded Green's function. According to Szeg\H{o}'s theorem, the long time asymptotic is determined by the logarithm of this function, which is not an analytic function. The imaginary part of $C_{\alpha}^f(\omega)$ crosses the zero line only at $\omega=0$ and the winding number is determined by the sign of the real part at this point. Note that we can not calculate the winding number via its strict mathematical definition; instead we use Nyquist (Cole-Cole) type plots that illustrate their topology along with the changes of the sign of the imaginary part for negative values of the real one. 

\begin{figure}[!tb]
	\centering
	\includegraphics[width=0.35\textwidth]{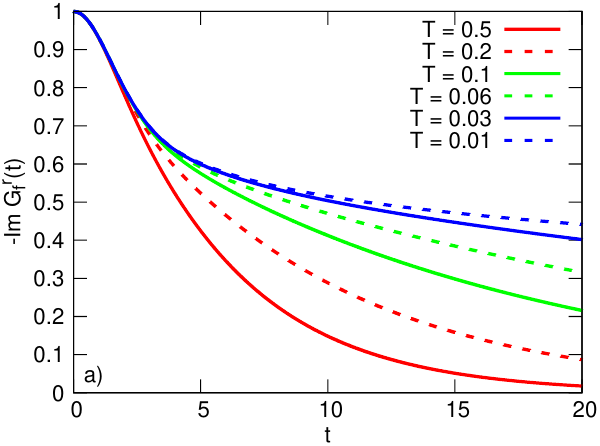} \\
	\includegraphics[width=0.35\textwidth]{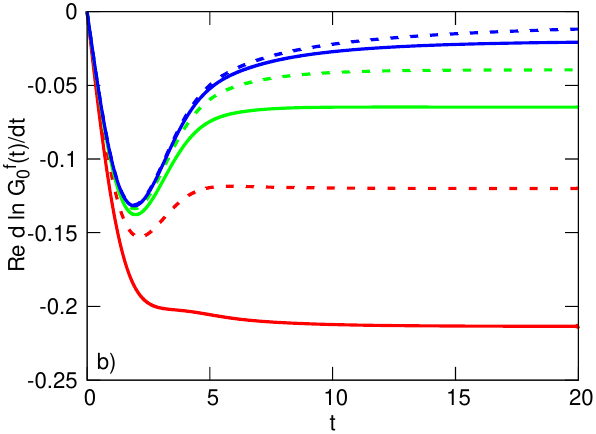} \\
	\includegraphics[width=0.35\textwidth]{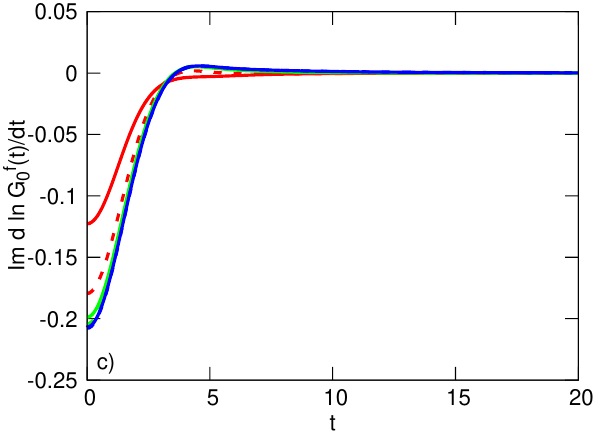} \\
	\includegraphics[width=0.35\textwidth]{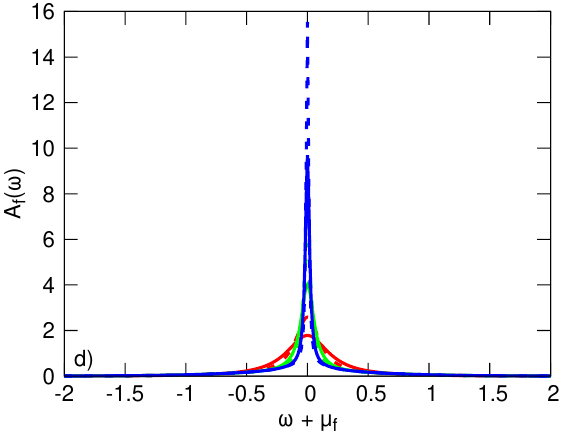} 
	\caption{(a) The $f$-electron propagator $G_f^{>}(t)$, (b,c) its logarithmic derivative, and (d) DOS $A_f(\omega)$ for $U=0.8$, $n_f=0.5$, $n_d=0.5$.}
\label{fig:fU08}
\end{figure}

At half filling and for $U<0.866$, when both $1- U \operatorname{Re} \mathcal{G}_{0}^t(\omega=0)>0$ and $1+ U \operatorname{Re} \mathcal{G}_{1}^t(\omega=0)>0$, the winding number is equal to zero (see Appendix~\ref{sec:fwinding}). Its logarithm is well defined and the corresponding determinants decay exponentially at large times for high temperatures, while they behave as a power-law for low temperatures (with a singular power law at $T=0$ only). This behavior can be seen in the change of the asymptotics of the logarithmic derivative from a constant to a $c/t$ dependence (Fig.~\ref{fig:fU08}). The corresponding DOS displays a Lorentzian peak for high temperatures, which is replaced by a power law singularity in the limit as $T\to 0$ in agreement with the seminal results of Brandt and Urbanek~\cite{brandt_1992:297}. 

\begin{figure}[!tb]
	\centering
	\includegraphics[width=0.35\textwidth]{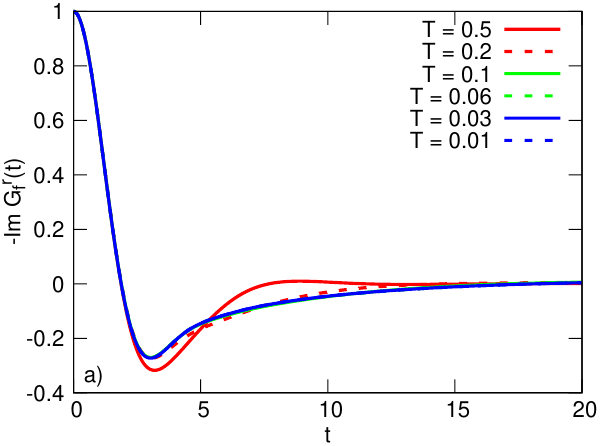} \\
	\includegraphics[width=0.35\textwidth]{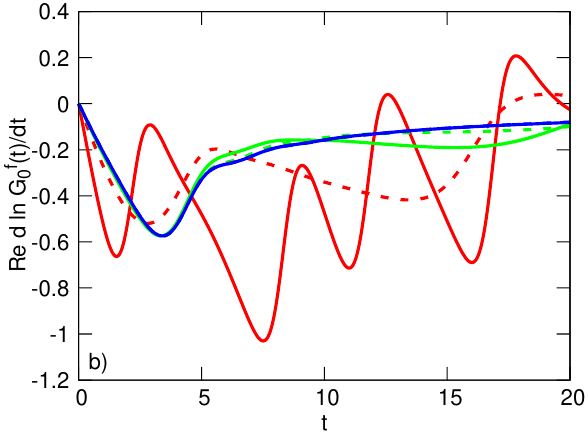} \\
	\includegraphics[width=0.35\textwidth]{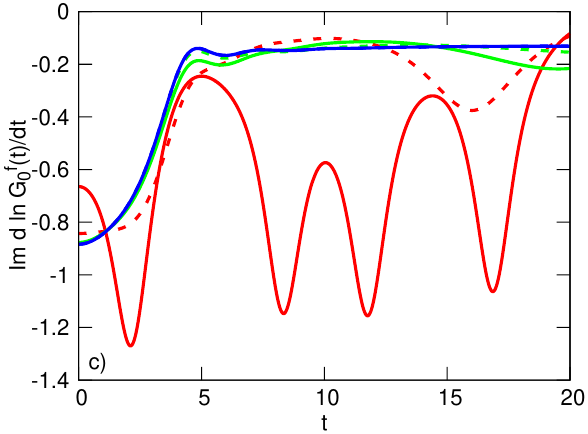} \\
	\includegraphics[width=0.35\textwidth]{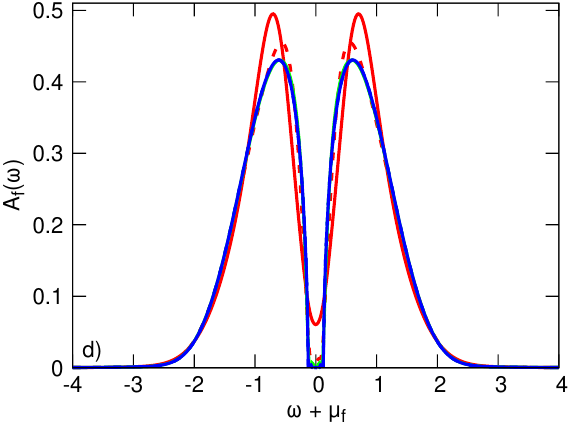}
	\caption{Same as in Fig.~\ref{fig:fU08} for $U = 2.0$.}
	\label{fig:fU20}
\end{figure}

On the other hand, for $U>0.866$, when both $1 \mp U \operatorname{Re} \mathcal{G}_{0,1}^t(\omega=0)<0$, the logarithm of the characteristic function is multivalued with a branch cut, and the winding number is equal to $-1$ (see Appendix~\ref{sec:fwinding}). In this case, the characteristic function \eqref{eq:Cfw} is replaced by 
\begin{equation}
\check{C}_{\alpha}^f(\omega) = -\mathrm{e}^{\mathrm{i}\omega W} C_{\alpha}^f(\omega),
\label{eq:Cbfw}
\end{equation}
where $W=\Delta t\to0$ is the time discretization, and the long time asymptotic becomes a more complicated nonexponential one with oscillations of the logarithmic derivatives (Fig.~\ref{fig:fU20}). Such oscillations are strongest at high temperatures and decrease their amplitude and increase their period as the temperature is lowered. Now, the corresponding DOS possesses a gap at low temperatures due to freezing of thermal fluctuations~\cite{brandt_1992:297} and the presence of strong-correlations creating a Mott gap; the gap region is filled in by thermal excitations for increasing temperatures.

Szeg\H{o}'s theorem and the Wiener-Hopf approach work better for smooth functions. In our case, the functions \eqref{eq:iG0} and \eqref{eq:iG1} are smooth at high temperatures, whereas at low temperatures, the Fermi distribution $f(\omega)$ becomes step-like and one must employ results for singular integral equations~\cite{nozieres_1969:1097,muskhelishvili_1977:SingularIntegral}. Thus, at $T=0$, the long time asymptotic becomes a power law in the metallic phase for $U<U_c=\sqrt{2}$ and rapidly goes to zero for the Mott insulator ($U>U_c$).

\section{Lattice gas}

As mentioned above, the FKM can be projected onto a classical lattice gas (Ising) model with effective interactions. So, we first consider the simple lattice gas model
\begin{equation}
	H = \frac{1}{2} \sum_{ij} V_{ij} n_i n_j - \mu \sum_i n_i,
\end{equation}
where $n_i=c_i^{\dagger}c_i$ is the occupation of site $i$ by a spinless Fermi-type particle, with values $n_i=1$ or $0$. This model corresponds to a classical Ising model with $S_i^z = n_i - \frac{1}{2} = \pm\frac{1}{2}$.

Despite the occupation operator $n_i$ 
being a conserved quantity, $[n_i,H]=0$, one can still examine nontrivial behavior for the dynamics and for the density of states for such fermions. From the equations of motion, we have that
\begin{align}
	\frac{\mathrm{d} c_i(t)}{\mathrm{d}t} &= -\mathrm{i}\Bigl[\sum_j V_{ij} n_j(t) - \mu\Bigr] c_i(t),
	\\
	\frac{\mathrm{d} c_i^{\dagger}(t)}{\mathrm{d}t} &= \mathrm{i}\Bigl[\sum_j V_{ij} n_j(t) - \mu\Bigr] c_i^{\dagger}(t).
\end{align}
This then yields the time evolution of the annihilation and creation operators, respectively,
\begin{align}
	c_i(t) &= U_i(t,0) c_i(0) = c_i(0) \bar{U}_i^{\dagger}(t,0),
	\\
	c_i^{\dagger}(t) &= \bar{U}_i(t,0) c_i^{\dagger}(0) = c_i^{\dagger}(0) U_i^{\dagger}(t,0),
\end{align}
where the evolution operators are given by
\begin{align}
	U_i(t,0) &= T\exp\biggl\{-\mathrm{i}\int_{0}^{t}\mathrm{d}t'\Bigl[\sum_j V_{ij} n_j(t') - \mu\Bigr]\biggr\},
	\\
	\bar{U}_i(t,0) &= T\exp\biggl\{\mathrm{i}\int_{0}^{t}\mathrm{d}t'\Bigl[\sum_j V_{ij} n_j(t') - \mu\Bigr]\biggr\}.
\end{align}
Now, for the single-particle Green's function we find [$G_{ij}(t,t') = \delta_{ij}G_i(t,t')$]
\begin{align}
	G_i^{>}(t,0) &= -\mathrm{i}\bigl\langle U_i(t,0)c_i(0)c_i^{\dagger}(0)\bigr\rangle 
	\nonumber\\
	&= -\mathrm{i}\left\langle \mathrm{e}^{-\mathrm{i}[\sum_{j}V_{ij}n_j-\mu]t}(1-n_i)\right\rangle ,
	\\
	G_i^{<}(t,0) &= \mathrm{i}\left\langle \mathrm{e}^{-\mathrm{i}[\sum_{j}V_{ij}n_j-\mu]t}n_i\right\rangle ,
	\\
	G_i^{r}(t,0) &= -\mathrm{i}\Theta(t)\left\langle \mathrm{e}^{-\mathrm{i}[\sum_{j}V_{ij}n_j-\mu]t}\right\rangle ,
	\label{eq:Gfr}
\end{align}
where we have used the fact that $n_i(t) = n_i$ is a conserved quantity, which allows us to replace the time-ordered exponents by ordinary ones. 

In the case of a nearest-neighbor interaction, one has
\begin{equation}\label{eq:V_Z}
	V_{ij} = \frac{V^{*}}{Z}, 
\end{equation} 
where $Z$ is the coordination number (number of nearest neighbors for site $i$). Calculating the local density of states, we obtain
\begin{align}
	A_i(\omega) &= \Bigl\langle \delta\Bigl(\omega+\mu-\sum_{j}V_{ij}n_j\Bigr)\Bigr\rangle
	\nonumber\\
	&= \sum_{s=0}^{Z} w_i(s) \delta\Bigl(\omega+\mu-V^{*}\frac{s}{Z}\Bigr).
\end{align}
Here $w_i(s)$ are statistical weights of the configurations with $s$ occupied neighbors of site $i$, and
\begin{equation}
	 \sum_{s=0}^Z w_i(s)=1. 
\end{equation}
Then, for the homogeneous phase, we have
\begin{equation}
	\sum_{s=0}^Z w_i(s) \frac{s}{Z} = \langle n_j\rangle = \langle n\rangle.
\end{equation}

Now, we examine the OTOCs for the lattice gas model. For the square modulus of the anticommutator, we have
\begin{align}
	&\bigl\langle\bigl|\bigl[c_i(t),c_j^{\dagger}(0)\bigr]_{+}\bigr|^2\bigr\rangle 
	\nonumber\\
	&= \bigl\langle c_j(0) c_i^{\dagger}(t) c_i(t) c_j^{\dagger}(0)\bigr\rangle
	+ \bigl\langle c_i^{\dagger}(t) c_j(0) c_j^{\dagger}(0) c_i(t)\bigr\rangle
	\nonumber\\
	&+ \bigl\langle c_j(0) c_i^{\dagger}(t) c_j^{\dagger}(0) c_i(t)\bigr\rangle
	+ \bigl\langle c_i^{\dagger}(t) c_j(0) c_i(t) c_j^{\dagger}(0)\bigr\rangle.
\end{align}
Using $c_i^{\dagger}(t) c_i(t) = n_i$, $c_j(0) c_j^{\dagger}(0) = 1-n_j$, and 
\begin{equation}
	\mathrm{e}^{\mathrm{i}[\sum_{l}V_{il}n_l-\mu]t} c_j^{\dagger} \mathrm{e}^{-\mathrm{i}[\sum_{l}V_{il}n_l-\mu]t} = \mathrm{e}^{i V_{ij} t} c_j^{\dagger},
\end{equation}
we find both local and nonlocal terms
\begin{align}
	&\bigl\langle\bigl|\bigl[c_i(t),c_j^{\dagger}(0)\bigr]_{+}\bigr|^2\bigr\rangle 
	\nonumber\\
	&= \delta_{ij} + (1-\delta_{ij})\, \langle n_i (1-n_j)\rangle\, 4\sin^2\frac{V_{ij} t}{2}. \label{eq:LGsqmod}
\end{align}
Similarly, another OTOC gives us
\begin{align}
	&\bigl\langle\bigl[c_i^{\dagger}(t),c_j^{\dagger}(0)\bigr]_{+}\bigl[c_i(t),c_j(0)\bigr]_{+}\bigr\rangle 
	\nonumber\\
	&= (1-\delta_{ij})\, \langle n_i n_j\rangle\, 4\sin^2\frac{V_{ij} t}{2},
	\\
	&\bigl\langle\bigl[c_i(t),c_j(0)\bigr]_{+}\bigl[c_i^{\dagger}(t),c_j^{\dagger}(0)\bigr]_{+}\bigr\rangle 
	\nonumber\\
	&= (1-\delta_{ij})\, \langle (1-n_i) (1-n_j)\rangle\, 4\sin^2\frac{V_{ij} t}{2},
\end{align}
which are also nonlocal. One can see, that these OTOCs expand only with the distance of the interaction range $V_{ij}$ and represent a direct instantaneous interaction. In the limit $Z\to\infty$ for the nearest-neighbor interactions \eqref{eq:V_Z}, the nonlocal contributions vanish as $1/Z$.

On the other hand, for the square of the anticommutator, we have only local contributions given by
\begin{equation}
	\bigl\langle\bigl[c_i(t),c_j^{\dagger}(0)\bigr]_{+}^2\bigr\rangle = \delta_{ij} \left\langle \mathrm{e}^{-2\mathrm{i}[\sum_l V_{il} n_l - \mu]t} \right\rangle,
	\label{eq:LG_OTOCsq}
\end{equation}
while for nearest-neighbor interactions \eqref{eq:V_Z}, we find
\begin{equation}
	\bigl\langle\bigl[c_i(t),c_j^{\dagger}(0)\bigr]_{+}^2\bigr\rangle = \delta_{ij} \sum_{s=0}^{Z} w_s \mathrm{e}^{-2\mathrm{i}[V^{*} \frac{s}{Z} - \mu]t} .
\end{equation}
One can see, that the expression~\eqref{eq:LG_OTOCsq} for the OTOC constructed by the square of the anticommutators is similar to the one for the retarded Green's function~\eqref{eq:Gfr}, both having the square of the exponent (evolution operator). It can potentially be considered as a higher moment of a (quasi)probability.

\section{OTOCs for the Falicov-Kimball model}

Similar to the lattice gas model, we can examine different OTOCs for the $f$ electrons of the FKM.

\subsection{Anticommutator squared magnitude}

\begin{figure}[!htb]
	\centering
	\includegraphics[width=0.7\linewidth]{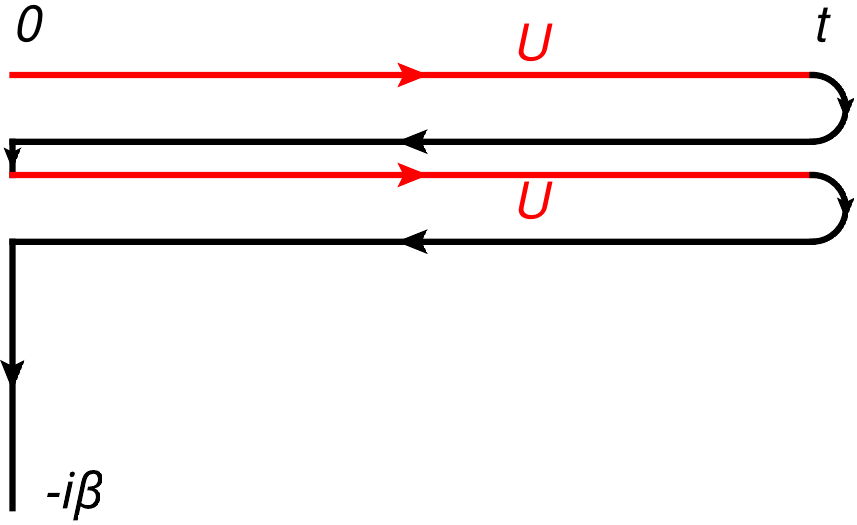}
	\caption{Doubly folded Keldysh contour. The interaction $U$ in~\eqref{eq:Udef} and \eqref{eq:Fsq_defU_cont} acts on the forward (upper) branches of the contour.}
	\label{fig:doublekeldysh}
\end{figure}

The first one is the squared magnitude of the anticommutator of the creation and annihilation operators at different times $[f_i(t),f_j^{\dagger}(0)]_{+}$, which satisfy
\begin{align}
&F_{ij}^{\textrm{mod}}(t) = \bigl\langle\bigl|[f_i(t),f_j^{\dagger}(0)]_{+}\bigr|^2\bigr\rangle
\nonumber\\
&= \bigl\langle[f_i(t),f_j^{\dagger}(0)]_{+}[f_i(t),f_j^{\dagger}(0)]_{+}^{\dagger}\bigr\rangle
\nonumber\\
&= \bigl\langle f_i(t) f_j^{\dagger}(0) f_j(0) f_i^{\dagger}(t)\bigr\rangle
 + \bigl\langle f_j^{\dagger}(0) f_i(t) f_i^{\dagger}(t) f_j(0)\bigr\rangle
\nonumber\\
&+ \bigl\langle f_i(t) f_j^{\dagger}(0) f_i^{\dagger}(t) f_j(0)\bigr\rangle
+ \bigl\langle f_j^{\dagger}(0) f_i(t) f_j(0) f_i^{\dagger}(t)\bigr\rangle
\nonumber\\
&= \bigl\langle f_i f_j^{\dagger} f_j f_i^{\dagger}\bigr\rangle
+ \bigl\langle f_j^{\dagger} f_i f_i^{\dagger} f_j\bigr\rangle
\nonumber\\
&+ 2 \operatorname{Re} \bigl\langle f_i(t) f_j^{\dagger}(0) f_i^{\dagger}(t) f_j(0)\bigr\rangle.
\label{eq:Fmod_def}
\end{align}
For the FKM, the product of the creation and annihilation operators at the same times in the first two terms give either $P_{i}^{+} = f_i^{\dagger}(t) f_i(t)$ or $P_{i}^{-} = f_i(t) f_i^{\dagger}(t)$, which are  conserved quantities, and do not depend on time. These terms yield static correlation functions. 
But, the last two terms of the time evolution of the operators are determined by \eqref{eq:f_t} and \eqref{eq:fd_t} and, as a result, we find
\begin{align}
F_{ij}^{\textrm{mod}}(t) &=
\bigl\langle f_i f_j^{\dagger} f_j f_i^{\dagger}\bigr\rangle
+ \bigl\langle f_j^{\dagger} f_i f_i^{\dagger} f_j\bigr\rangle
\nonumber\\
&+ 2\operatorname{Re}\bigl\langle \mathcal{U}_i(t,0) f_i f_j^{\dagger} f_i^{\dagger} \mathcal{U}_i^{\dagger}(t,0) f_j\bigr\rangle,
\label{eq:Fmod_defU}
\end{align}
where in the last term we have an evolution on the doubly folded Keldysh contour (Fig.~\ref{fig:doublekeldysh}).

For the local function, $i=j$, we have
\begin{equation}
F_{ii}^{\textrm{mod}}(t) = 1.
\end{equation}
The nonlocal function $(i\neq j)$ satisfies
\begin{align}
F_{ij}^{\textrm{mod}}(t) &=
2\bigl\langle  P_i^{-} P_j^{+}\bigr\rangle
\nonumber\\
&+ 2\operatorname{Re}\bigl\langle \mathcal{U}_i(t,0) f_i f_j^{\dagger} f_i^{\dagger} \mathcal{U}_i^{\dagger}(t,0) f_j\bigr\rangle
\nonumber \\
&=
2\bigl\langle  P_i^{-} P_j^{+}\bigr\rangle
\nonumber\\
&- 2\operatorname{Re}\bigl\langle P_i^{-} \mathcal{U}_i(t,0) f_j^{\dagger} \mathcal{U}_i^{\dagger}(t,0) f_j\bigr\rangle.
\label{eq:Fmod_nonloc}
\end{align}
In the infinite-dimensional case, it contains two evolution operators \eqref{eq:Udef} which act on the site $i$ and do not touch the site $j$. As a result, we can carry one of the evolution operators through the $f_j^{\dagger}$ and they mutually cancel. Finally, we obtain
\begin{align}
F_{ij}^{\textrm{mod}}(t) &= \delta_{ij}.
\label{eq:Fmod_nonloc_fin}
\end{align}
On the other hand, in the finite $D\to\infty$ case, one has to replace the evolution operators \eqref{eq:Udef} and \eqref{eq:tUdef} for the local impurity problem by the evolution operators for the lattice, which makes it much more difficult to find a solution and we do not consider it further here. This does give nonlocal $1/Z$-corrections, but, in contrast to \eqref{eq:LGsqmod}, they are governed by the effective long-range retarded interactions.

\subsection{Other trivial anticommutator functions}

We can also consider an anticommutator of two creation $[f_i^{\dagger}(t),f_j^{\dagger}(0)]_{+}$ or two annihilation $[f_i(t),f_j(0)]_{+}$ operators and construct another OTOC. The square of such anticommutators yield OTOC functions which are equal to zero. We have
\begin{align}
&F_{ij}^{\textrm{sq},c}(t) = \bigl\langle[f_i^{\dagger}(t),f_j^{\dagger}(0)]_{+}[f_i^{\dagger}(t),f_j^{\dagger}(0)]_{+}\bigr\rangle =0,
\nonumber\\
&F_{ij}^{\textrm{sq},a}(t) = \bigl\langle[f_i(t),f_j(0)]_{+}[f_i(t),f_j(0)]_{+}\bigr\rangle =0.
\label{eq:Fsq_cc_def}
\end{align}
For the squared magnitude, similar to \eqref{eq:Fmod_nonloc}, we find that
\begin{align}
&F_{ij}^{\textrm{mod},ca}(t) = \bigl\langle[f_i(t),f_j(0)]_{+}[f_i(t),f_j(0)]_{+}^{\dagger}\bigr\rangle
\nonumber\\
&= \bigl\langle[f_i(t),f_j(0)]_{+}[f_i^{\dagger}(t),f_j^{\dagger}(0)]_{+}\bigr\rangle
\nonumber\\
&= (1-\delta_{ij}) 2\bigl\langle P_i^{-} P_j^{-} \bigr\rangle
\nonumber\\
&+ (1-\delta_{ij}) 2 \operatorname{Re} \bigl\langle \mathcal{U}_i(t,0) f_i f_j f_i^{\dagger} \mathcal{U}_i^{\dagger}(t,0) f_j^{\dagger}\bigr\rangle =0.
\label{eq:Fmod_cc_def}
\end{align}
~

\section{Square of the anticommutator}

For the square of anticommutator, we have
\begin{align}
&F_{ij}^{\textrm{sq}}(t) = \bigl\langle\bigl([f_i(t),f_j^{\dagger}(0)]_{+}\bigr)^2\bigr\rangle
\nonumber\\
&= \bigl\langle[f_i(t),f_j^{\dagger}(0)]_{+}[f_i(t),f_j^{\dagger}(0)]_{+}\bigr\rangle 
\nonumber\\
&= \bigl\langle f_i(t) f_j^{\dagger}(0) f_i(t) f_j^{\dagger}(0)\bigr\rangle
+ \bigl\langle f_j^{\dagger}(0) f_i(t) f_j^{\dagger}(0) f_i(t)\bigr\rangle
\nonumber\\
&+ \bigl\langle f_i(t) f_j^{\dagger}(0) f_j^{\dagger}(0) f_i(t)\bigr\rangle
+ \bigl\langle f_j^{\dagger}(0) f_i(t) f_i(t) f_j^{\dagger}(0)\bigr\rangle.
\label{eq:Fsq_def}
\end{align}
The last two terms are equal to zero and we obtain
\begin{align}
F_{ij}^{\textrm{sq}}(t) &= \bigl\langle \mathcal{U}_i(t,0) f_i f_j^{\dagger} \mathcal{U}_i(t,0) f_i f_j^{\dagger}\bigr\rangle
\nonumber\\
&+ \bigl\langle \tilde{\mathcal{U}}_i(t,0) f_i^{\dagger} f_j \tilde{\mathcal{U}}_i(t,0) f_i^{\dagger} f_j\bigr\rangle^{*}.
\label{eq:Fsq_defU}
\end{align}
The evolution operators do not change the occupation of the $f$ states and the nonlocal contributions with $i\neq j$ contain products of two creation and two annihilation operators at the same sites, which vanish. Hence, in the $D=\infty$ limit, we have only local OTOCs that are nonzero. They are given by
\begin{align}
	F_{ij}^{\textrm{sq}}(t) &= \delta_{ij} F_{ii}^{\textrm{sq}}(t),
	\\
F_{ii}^{\textrm{sq}}(t) &= \bigl\langle \mathcal{U}_i(t,0) \mathcal{U}_i(t,0) P_i^{-} \bigr\rangle
\nonumber\\
&+ \bigl\langle \tilde{\mathcal{U}}_i(t,0) \tilde{\mathcal{U}}_i(t,0) P_i^{+} \bigr\rangle^{*}.
\label{eq:Fsq_defU_loc}
\end{align}
This expression can be rewritten in a form similar to the one for the $f$-electron propagator \cite{shvaika_2008:425}
\begin{widetext}
\begin{align}
&F_{ii}^{\textrm{sq}}(t) = \frac{1}{Z} \mathrm{e}^{-\mathrm{i}2(E_f-\mu)t} \operatorname{Tr} \biggl\{\mathrm{e}^{\beta\mu n_{id}} \mathcal{T}_{c}\exp\left[-\mathrm{i}\!\int\nolimits_{c}\!\mathrm{d}t'\!\int\nolimits_{c}\!\mathrm{d}t' d_i^{\dagger}(t')\lambda_c(t',t'')d_i(t'') - \mathrm{i}\!\int\nolimits_{c}\!\mathrm{d}t'U_c^{[0,t],[0,t]}(t') n_{id}(t')\right] \biggr\}
\nonumber\\
&+ \frac{\mathrm{e}^{\beta(\mu-E_f)}}{Z} \mathrm{e}^{-\mathrm{i}2(E_f-\mu)t} \left(\operatorname{Tr} \biggl\{\mathrm{e}^{\beta(\mu-U) n_{id}} \mathcal{T}_{c}\exp\left[-\mathrm{i}\!\int\nolimits_{c}\!\mathrm{d}t'\!\int\nolimits_{c}\!\mathrm{d}t' d_i^{\dagger}(t')\lambda_c(t',t'')d_i(t'') +\mathrm{i}\!\int\nolimits_{c}\!\mathrm{d}t'U_c^{[0,t],[0,t]}(t') n_{id}(t')\right] \biggr\}\right)^{*},
\label{eq:Fsq_defU_cont}
\end{align}
where the integration and time ordering are performed over the doubly folded Keldysh contour (Fig.~\ref{fig:doublekeldysh}). Now the time-dependent field $U_c^{[0,t],[0,t]}(t')$ is nonzero and equal to $U$ on the two forward branches for $t'\in[0,t]$ and is zero otherwise. In the same way as in \cite{shvaika_2008:425}, we can derive the final expression for the OTOC:
\begin{align}
F_{ii}^{\textrm{sq}}(t) &= \mathrm{e}^{-\mathrm{i}2(E_f-\mu)t} F(t),
\nonumber\\
F(t) &= F_0(t) + F_1^{*}(t),
\label{eq:Ft_fin}\\
F_0(t) &= w_0 \det\nolimits_{[0,t],[0,t]} \left[ \delta_c(t',t'') - G_0 (t',t'') U\right],
\nonumber\\
F_1(t) &= w_1 \det\nolimits_{[0,t],[0,t]} \left[ \delta_c(t',t'') + G_1 (t',t'') U\right].
\label{eq:F01t_fin}
\end{align}
In a discretized form, $t_m = m W $ with $t_0=0$ and $t_N=t$, the first determinant looks like
\begin{align}
&\det\nolimits_{[0,t],[0,t]}\left[\begin{array}{c|c}
\mathbf{I} - W\mathcal{G}_{0}^t U & - W\mathcal{G}_{0}^{<} U \\
\hline
- W\mathcal{G}_{0}^{>} U & \mathbf{I} - W\mathcal{G}_{0}^t U 
\end{array}\right]
= \det\nolimits_{[0,t]}\left[\mathbf{I} - W\mathcal{G}_{0}^t U\right] \det\nolimits_{[0,t]}\left[\mathbf{I} - W\mathcal{G}_{0}^t U - W\mathcal{G}_{0}^{<} U \left(\mathbf{I} - W\mathcal{G}_{0}^t U\right)^{-1} W\mathcal{G}_{0}^{>} U\right]
\nonumber \\
&=\det\left[\begin{array}{ccc|ccc}
1 - W \mathcal{G}_0^t(0) U & \cdots & - W \mathcal{G}_0^{<}(-t) U & - W \mathcal{G}_0^{<}(0) U & \cdots & - W \mathcal{G}_0^{<}(-t) U \\
\vdots & \ddots & \vdots & \vdots & \ddots & \vdots \\
- W \mathcal{G}_0^{>}(t) U & \cdots & 1 - W \mathcal{G}_0^t(0) U & - W \mathcal{G}_0^{<}(t) U & \cdots & - W \mathcal{G}_0^{<}(0) U \\
\hline
- W \mathcal{G}_0^{>}(0) U & \cdots & - W \mathcal{G}_0^{>}(-t) U & 1 - W \mathcal{G}_0^t(0) U & \cdots & - W \mathcal{G}_0^{<}(-t) U \\
\vdots & \ddots & \vdots & \vdots & \ddots & \vdots \\
- W \mathcal{G}_0^{>}(t) U & \cdots & - W \mathcal{G}_0^{>}(0) U  & - W \mathcal{G}_0^{>}(t) U & \cdots & 1 - W \mathcal{G}_0^t(0) U 
\end{array}\right],
\label{eq:OTOCdet}
\end{align}
\end{widetext}
where
\begin{align}
\mathcal{G}_0^{<}(t'-t'') &= - \frac{\mathrm{i}}{\pi} \int_{-\infty}^{+\infty}\mathrm{d}\omega\, \mathrm{e}^{-\mathrm{i}\omega(t'-t'')} f(\omega) 
\nonumber \\
&\times\operatorname{Im} \frac{1}{\omega + \mu -\lambda(\omega)}
\label{eq:G0l}
\end{align}
is the lesser Green's function,
\begin{align}
\mathcal{G}_0^{>}(t'-t'') &= - \frac{\mathrm{i}}{\pi} \int_{-\infty}^{+\infty}\mathrm{d}\omega\, \mathrm{e}^{-\mathrm{i}\omega(t'-t'')} \left[f(\omega)-1\right]  
\nonumber \\
&\times\operatorname{Im} \frac{1}{\omega + \mu -\lambda(\omega)}
\label{eq:G0g}
\end{align}
is the greater one, and for the equal times $t'=t''$ we choose $\Theta(0)=0$~\cite{pakhira_2019:125137}, which gives
\begin{align}
\mathcal{G}_0^t(0) &= \mathcal{G}_0^{<}(0) = - \frac{\mathrm{i}}{\pi} \int_{-\infty}^{+\infty}\mathrm{d}\omega\, f(\omega)  
\nonumber \\
&\times\operatorname{Im} \frac{1}{\omega + \mu -\lambda(\omega)}.
\label{eq:G0c}
\end{align}
Similar expressions can be written for the second determinant in~\eqref{eq:Ft_fin}.

One might expect these results to be similar to those of the $f$-particle propagator. Indeed, they are. We find that the long time asymptotic of the determinants in~\eqref{eq:OTOCdet} or $F_{\alpha}(t)$ in \eqref{eq:F01t_fin} are determined by the Szeg\H{o}-Widom theorem for block Toeplitz matrices~\cite{widom_1976:1}, which employ characteristic functions
\begin{align}
C_{0,1}(\omega) &= \left[1 \mp U \mathcal{G}_{0,1}^t(\omega)\right]^2 - U^2 \mathcal{G}_{0,1}^{>}(\omega) \mathcal{G}_{0,1}^{<}(\omega)
\nonumber \\
&= \left[1 \mp U \operatorname{Re} \mathcal{G}_{0,1}^r(\omega)\right]^2 -U^2 \left[\operatorname{Im}\mathcal{G}_{0,1}^r(\omega)\right]^2
\nonumber \\
&\mp 2\mathrm{i} U \tanh\frac{\beta\omega}{2}\operatorname{Im}\mathcal{G}_{0,1}^r(\omega)\left[1 \mp U \operatorname{Re} \mathcal{G}_{0,1}^r(\omega)\right].
\label{eq:Cw}
\end{align}
Now, the imaginary part of each characteristic function can become zero either at one frequency, $\omega=0$, or at two frequencies, $\omega=0$ and when $1 \mp U \operatorname{Re} \mathcal{G}_{0,1}^r(\omega)=0$. We examine this OTOC in detail in the next section.

\section{Results and Discussion}

\begin{figure}[!tb]
	\centering
	\includegraphics[width=0.35\textwidth]{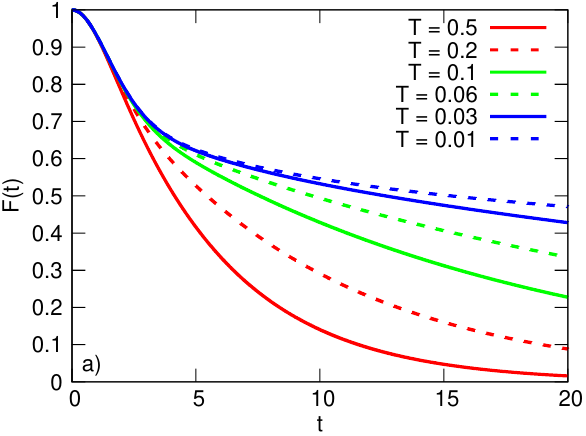} \\
	\includegraphics[width=0.35\textwidth]{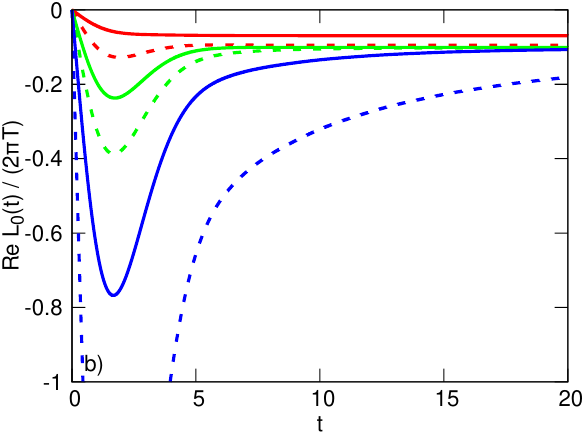} \\
	\includegraphics[width=0.35\textwidth]{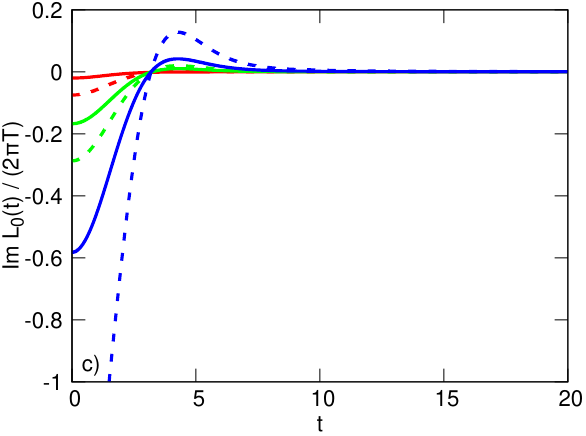} \\ 
	\includegraphics[width=0.35\textwidth]{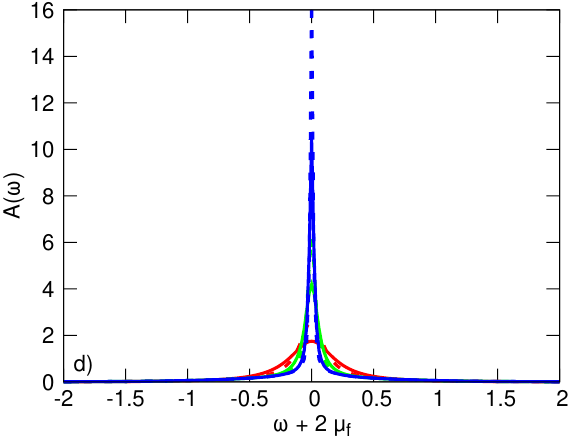}
	\caption{(a) OTOC $F(t)$, (b,c) normalized logarithmic derivative $L_0(t)/(2\pi T)$, and (d) the corresponding ``density-of-states'' $A(\omega)$ for $U=0.4$, $n_f=0.5$, $n_d=0.5$.}
	\label{fig:U04}
\end{figure}

At half filling $n_d=n_f=0.5$, the contributions $F_0(t)$ and $F_1(t)$ in~\eqref{eq:F01t_fin} are equal and the total OTOC $F(t)$ is real. But, the partial contributions from the sites with empty $f$ particle states, $F_0(t)$, and with occupied $f$ particle states, $F_1(t)$, display intriguing time dependencies.
For $U<0.44$, the real part of both characteristic functions~\eqref{eq:Cw} is positive at frequencies where the imaginary part vanishes and the winding number is zero (see Appendix~\ref{sec:OTOCwinding}). The OTOC function \eqref{eq:Ft_fin} displays exponential decay at large times for high temperatures, which is replaced by a power law as $T\to0$ (Fig.~\ref{fig:U04}). The corresponding logarithmic derivative
\begin{equation}
	L_0(t) = \frac{\mathrm{d} \ln F_0(t)}{\mathrm{d}t} = \frac{1}{F_0(t)} \frac{\mathrm{d} F_0(t)}{\mathrm{d}t}
	\label{eq:Lyapunov}
\end{equation}
approaches a constant value at large times for high temperatures, with a decay rate within the ``chaos bound'' $|L_0(t)|/(2\pi T)<1$~\footnote{Despite the chaos bound of \cite{maldacena_2016:106,xu_2019:031048} is applicable for the short timescale, we compare the decay rate with this temperature bound to elucidate thermal effects in relaxation.}, whereas it displays $1/t$ dependence for low temperatures, which confirms the power law asymptotics of the OTOC. Consequently,  the corresponding ``density-of-states''~\footnote{Now, DOS is defined by the real part of OTOC because \eqref{eq:Fsq_def} omits factor ``$-\mathrm{i}$'' in contrast to the one-particle Green's function \eqref{eq:Gfc_def}}
\begin{equation}\label{eq:OTOCdos}
A(\omega) = \frac{1}{\pi} \operatorname{Re} \int_{0}^{+\infty} \mathrm{d}t\; \mathrm{e}^{\mathrm{i}\omega t} F_{ii}^{\textrm{sq}}(t)
\end{equation}
displays a Lorentzian peak at high temperatures which is replaced by a power law singularity as $T\to 0$.

\begin{figure}[!tb]
	\centering
	\includegraphics[width=0.35\textwidth]{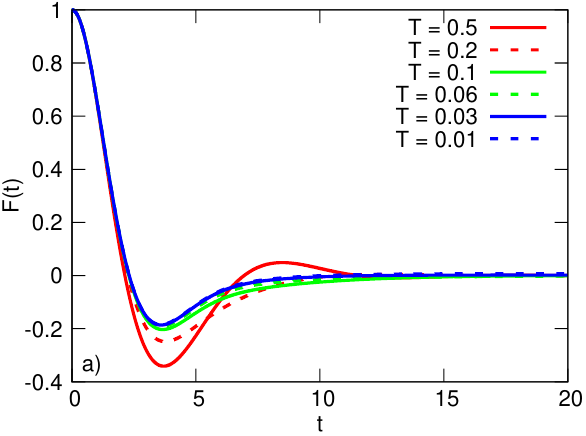} \\
	\includegraphics[width=0.35\textwidth]{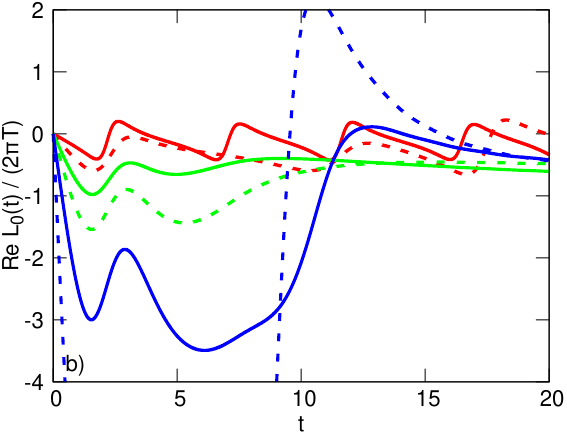} \\
	\includegraphics[width=0.35\textwidth]{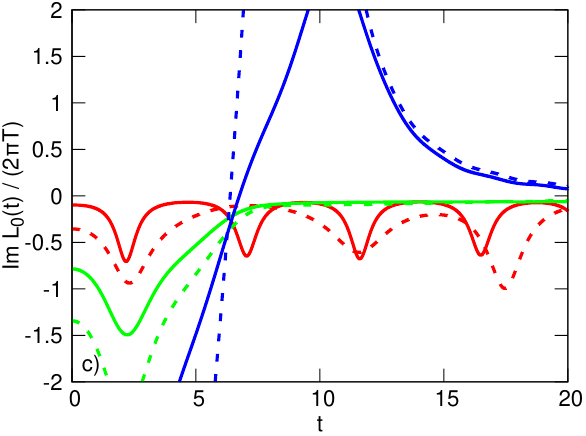} \\
	\includegraphics[width=0.35\textwidth]{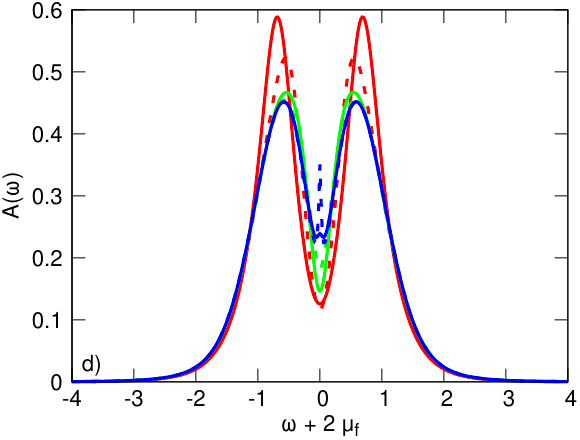}
	\caption{Same as in Fig.~\ref{fig:U04} for $U = 0.9$.}
	\label{fig:U09}
\end{figure}

Next, for $0.44<U<1.235$, the real part of the characteristic function becomes negative at $\omega=0$, the logarithm of the characteristic function is now a multivalued function with a branch cut and the winding number is equal to $-1$ (see Appendix~\ref{sec:OTOCwinding}). Now, the characteristic function \eqref{eq:Cw} must be replaced by 
\begin{equation}
\check{C}_{\alpha}(\omega) = -\mathrm{e}^{\mathrm{i}\omega W} C_{\alpha}(\omega),
\label{eq:Cbw}
\end{equation}
where $W=\Delta t\to0$ is the time discretization, and the long-time asymptotic becomes more complicated (Fig.~\ref{fig:U09}) with oscillations of the logarithmic derivatives $L_0(t)$ at high temperatures similar to the one for the single-particle propagators for $U>0.866$ (Fig.~\ref{fig:fU20}). With decreasing temperature, these oscillations are replaced by $1/t$ asymptotics, because we are still in the metallic phase $U<\sqrt2$. Now the OTOC's ``density-of-states'' displays a two peak structure separated by a pseudogap at high temperatures with the development of the power law singularity in the center of the pseudogap as $T\to0$. The sharp peak in the ``density-of-states'' is observed at low temperatures up to $U$ values where the ``density-of-states'' becomes negative, which reflects an increase in the quantum effects for these cases (such as Kirkwood-Dirac quasiprobabilities (KDQ)~\cite{yungerhalpern_2017:012120,yungerhalpern_2018:042105,gonzalezalonso_2019:040404}, which can also become negative).

\begin{figure}[!tb]
	\centering
	\includegraphics[width=0.35\textwidth]{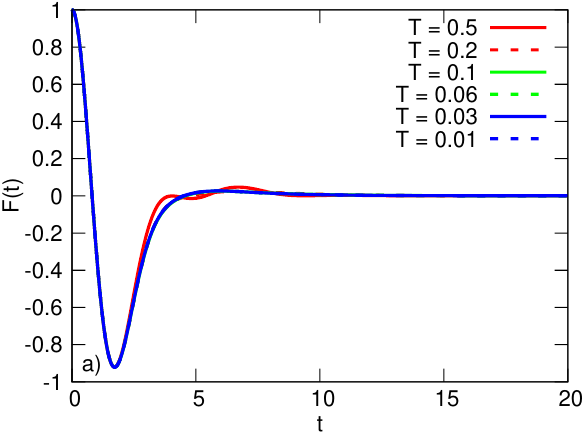} \\
	\includegraphics[width=0.35\textwidth]{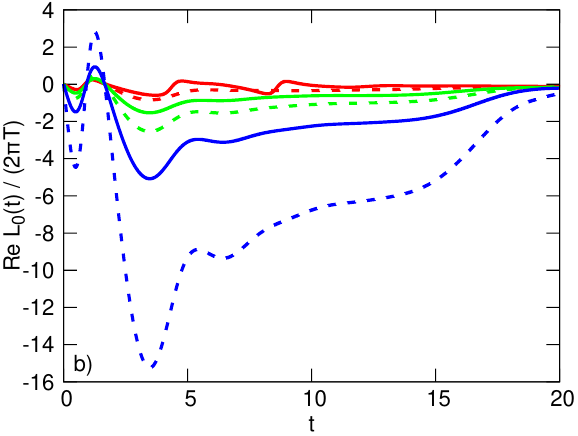} \\
	\includegraphics[width=0.35\textwidth]{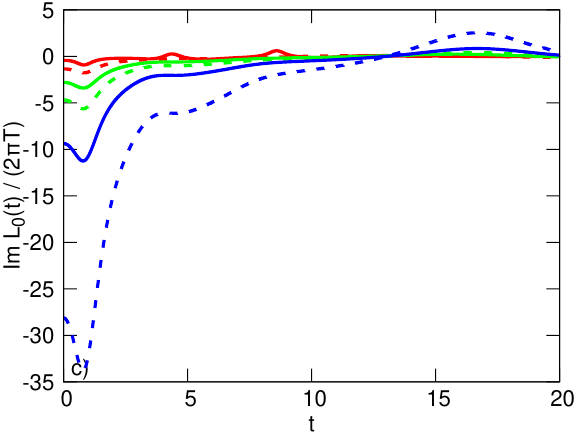} \\ 
	\includegraphics[width=0.35\textwidth]{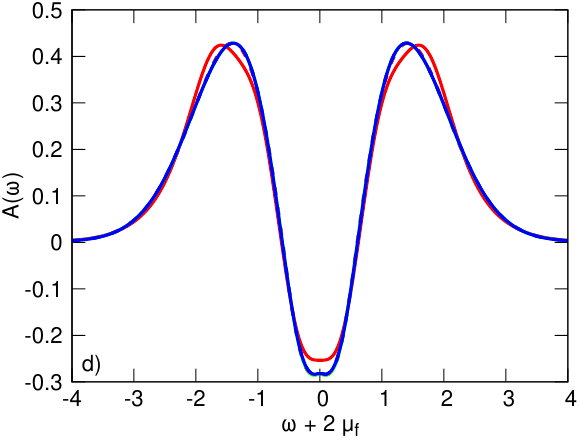}
	\caption{Same as in Fig.~\ref{fig:U04} for $U = 2.0$.}
	\label{fig:U20}
\end{figure}

For $U>0.866$, the imaginary part of the characteristic function \eqref{eq:Cw} crosses the zero line at frequencies where $1 \mp U \operatorname{Re} \mathcal{G}_{0,1}^r(\omega)=0$, but for $U<1.235$ the winding does not change because the real part is still positive at these points, whereas for $U>1.235$ the real part becomes negative and the winding number is equal to $-2$ (see Appendix~\ref{sec:OTOCwinding}) resulting in a more complicated long time behavior of the OTOCs and their logarithmic derivatives with damped oscillations (Fig.~\ref{fig:U20}). Now, the OTOC's DOS is more robust, displays weak temperature dependence, and possesses negative values in a wide frequency interval. For large values of interaction $U$, the quantum correlations become dominant and the ``flip'' of an $f$-state produces strong disturbance for the itinerant $d$-electrons, which can subsequently lead to local heat fluctuations. Such quantum heat fluctuations can be associated with negativity of quasiprobability distribution, e.g. Kirkwood-Dirac quasiprobability, which, on the other hand, can be a signature of suppressing of the decoherence and of quantum chaos and scrambling~\cite{levy_2020:010309,gherardini_2024:030201}.  

However, the sum rule (integral over frequency) for OTOC's DOS is conserved and is equal to one:
\begin{equation}
    \int_{-\infty}^{+\infty} \mathrm{d}\,\omega A(\omega) = F_{ii}^{\textrm{sq}}(0) = 1,
\end{equation}
which means that the appearance of negative DOS at small frequency values leads to a DOS enhancement elsewhere. For the first moment we obtain
\begin{equation}
    \int_{-\infty}^{+\infty} \mathrm{d}\,\omega\,\omega A(\omega) = \mathrm{i} \frac{\mathrm{d} F_{ii}^{\textrm{sq}}(t)}{\mathrm{d} t}\bigg|_{t=0} = 2 (U n_d + E_f -\mu),
\end{equation}
which equals zero at half filling $n_d=n_f=0.5$, and one can proceed similarly and systematically to determine higher-order moments.

It should be noted, that the Szeg\H{o}-Widom theorem for block Toeplitz matrices~\cite{widom_1976:1} is applicable for smooth functions. For this case in the fermionic FKM, the functions are continuous at high temperatures. At low temperatures, both the matrix elements in \eqref{eq:OTOCdet} and characteristic functions \eqref{eq:Cw} develop step-like features from the Fermi factors and the OTOC's long time behavior is now governed by localization singularities given by the Anderson orthogonality catastrophe~\cite{anderson_1967:1049,nozieres_1969:1097,pakhira_2019:125137}. These cases of three different winding numbers for the characteristic function are also distinguished by the behavior of the logarithmic derivative at low temperatures (see Appendix~\ref{sec:Lyapunov}, Fig.~\ref{fig:Lyapunov}).

At high temperatures, the oscillatory behavior of the logarithmic derivative is similar to the one seen in~\cite{lin_2018:144304,riddell_2019:054205} for  quantum spin models behind the light cone after the wave front passes. This analogy is plausible, because we consider a local OTOC function, which is always behind the wave front and at high temperature the effects of the Fermi statistics become negligible and the electrons behave like Pauli particles. Nevertheless, the localized $f$ particles act as  impurity scattering centers for the itinerant $d$ electrons and the destruction of the oscillations of logarithmic derivative for large Coulomb interaction $U$ is similar to the one observed in~\cite{riddell_2019:054205} with the an increase of disorder, when one observes the post wavefront ripples damped by disorder. This static disorder can protect the fast scrambling, which can be thought of as a breaking of the ``bound on chaos'', when at long times we have $|L_0(t)| > 2\pi k_B T/\hbar$, which, is also manifested by the negative OTOC ``density-of-states'' $A(\omega)$, see Fig.~\ref{fig:U20}. According to Yunger Halpern et al.~\cite{yungerhalpern_2018:042105}, OTOCs can be considered as moments of the Kirkwood-Dirac quasiprobabilities which implies that negative values of the ``density-of-states'' $A(\omega)$ are signatures of quantum information scrambling.

In cases where the calculated OTOC function decays with time, the question arises ``where has the corresponding quantum information gone?'' For lattice models with direct interactions between spins, it spreads to the nearby sites due to quantum information scrambling. For the FKM, there is no direct interaction between $f$ particles, only an effective retarded one mediated through the itinerant $d$ electrons, and in addition to quantum scrambling, quantum information can be transferred to the $d$ electrons. Exactly how this occurs is a project for future investigations.

\section{Conclusions}

We have examined how an indirect retarded interaction between qubits can affect the scrambling of quantum information. We considered the dynamics of the localized two-level states ($f$-electrons) interacting with the itinerant $d$-electrons in the Falicov-Kimball model and found exact results for the OTOCs constructed via state-flip operators ($f$-electron creation and annihilation operators) within dynamical mean field theory. Despite the classical (Ising) nature of our two-level states (the occupation $n_f$ is a conserved quantity), their dynamics is nontrivial and is governed by the winding of the characteristic functions for the corresponding Toeplitz matrices. With an increase of the local interaction potential $U$ between $f$ and $d$ electrons, the logarithmic derivative of the OTOCs start to display oscillations at high temperatures similar to those seen in quantum spin models. On the other hand, at low temperatures Fermi statistics and screening effects related to the orthogonality catastrophe become important, which strongly affects the OTOC's behavior. With a further increase of the interaction $U$, we find a fast damping of these oscillations due to the effects of disorder and, on the other hand, makes the OTOC's ``density-of-states'' (moments of the Kirkwood-Dirac quasiprobabilities) become negative, which is another manifestation of the quantum information scrambling.

Our results were obtained at half filling with $n_d=n_f=0.5$, but our main conclusions remain valid in when $n_f + n_d = 1$ holds, with the remark that now asymptotic behavior of each contribution $F_0(t)$ and $F_1(t)$ in \eqref{eq:Ft_fin} will be different with different windings of the corresponding Toeplitz matrices. On the other hand, in the doped case, with $n_f+n_d\ne 1$, our system will be in a metallic phase and the low temperature behavior will be dominated by the localization singularities given by the Anderson orthogonality catastrophe, whereas at high temperatures the results will be similar to the one for un-doped case.

\acknowledgments

This work was supported by the Department of Energy, Office of Basic Energy Sciences, Division of Materials Sciences and Engineering, under Grant No. DE-FG02-08ER46542.
J.K.F. was also supported by the McDevitt bequest.

\section*{Data Availability}

The data that support the findings of this article are openly available~\cite{shvaika_2025:OTOCFunctions}.


\appendix

\setcounter{figure}{0}
\renewcommand{\thefigure}{A.\arabic{figure}}

\section{Winding of characteristic function for one-particle propagator}
\label{sec:fwinding}

According to the Szeg\H{o}'s theorem (see~\cite{shvaika_2008:425} and references therein), the long time asymptotic of the $N\times N$ Toeplitz determinant
\begin{equation}
	D_N=\left | 
	\begin{array}{ccccc}
		c_0 & c_{-1} & c_{-2} & \hdots & c_{-N+1}\\
		c_1 & c_0 & c_{-1}& \hdots & c_{-N+2}\\
		c_2 & c_1 & c_0 & \hdots & c_{-N+3}\\
		\vdots & \vdots & \vdots & \vdots & \vdots\\
		c_{N-2} & c_{N-3} & c_{N-4} & \hdots & c_{-1} \\
		c_{N-1} & c_{N-2} & c_{N-3}& \hdots & c_0
	\end{array}
	\right |
\end{equation}
is equal to
\begin{equation}
	\lim_{N\rightarrow\infty} D_N=\exp\Bigl [ N g_0 + \sum_{n=1}^\infty n g_n g_{-n}\Bigr ]
	\label{eq: szego}
\end{equation}
with
\begin{equation}
	g_n=\frac{1}{2\pi}\int_{-\pi}^{\pi}\mathrm{d}\theta\, \mathrm{e}^{-\mathrm{i} n\theta}\ln C(\mathrm{e}^{\mathrm{i}\theta}).
	\label{eq: g_t_def}
\end{equation}
Here, $C(\xi)=\sum_{n=-\infty}^{+\infty}c_n\xi^n$ with $\xi=\mathrm{e}^{\mathrm{i}\theta}$ lying on the unit circle. The requirement for $C(\xi)$ to be continuous on the unit circle, is that its winding number (or index) is equal zero, namely
\begin{equation}
	\operatorname{Ind} C(\xi)=\frac{1}{2\mathrm{i}\pi}\left [ \ln C(\mathrm{e}^{2\mathrm{i}\pi})-\ln C(\mathrm{e}^{\mathrm{i}0})\right ] = 0,
	\label{eq: index_def}
\end{equation}
and we also require that the summation is well defined, or $\sum_{n=-\infty}^{\infty}|c_n|<\infty$ for the theorem to hold.

\begin{figure}[!b]
	\centering
	\includegraphics[width=0.35\textwidth]{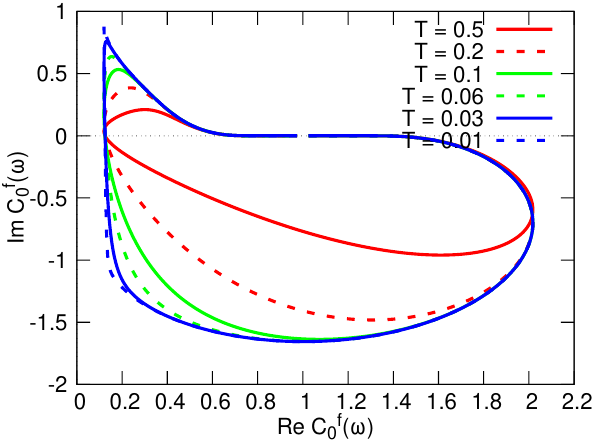}
	\caption{Winding of the characteristic function $C_{0}^f(\omega)$ for $U=0.8$, $n_f=0.5$, $n_d=0.5$.}
	\label{fig:fwindingU08}
\end{figure}

\begin{figure}[!tb]
	\centering
	\includegraphics[width=0.35\textwidth]{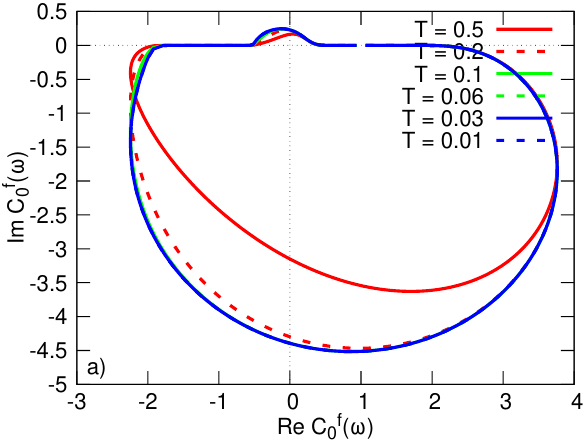} \\
	\includegraphics[width=0.35\textwidth]{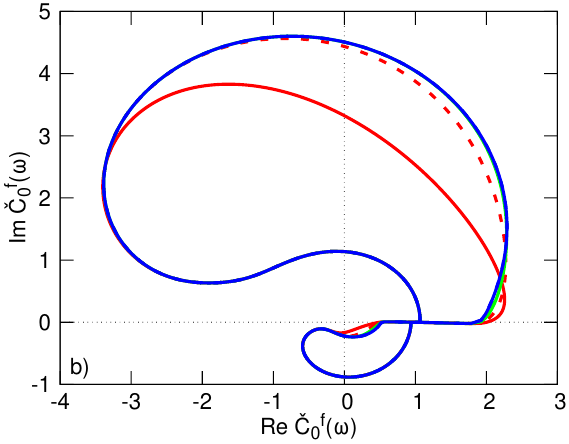}
	\caption{Winding of the characteristic functions (a) $C_{0}^f(\omega)$ and (b) $\check{C}_{0}^f(\omega)$ for $U=2.0$, $n_f=0.5$, $n_d=0.5$, and $\Delta t = 0.1$.}
	\label{fig:fwindingU10}
\end{figure}

After discretization of the time interval $[0,t]$ by $t_n=n\Delta t$, with $\Delta t=t/N$, the matrix elements $c_n$ in Eq.~\eqref{eq:Gfgr_def} satisfy
\begin{equation}
	c_n=\delta_{n0}-\Delta t U \mathcal{G}_0(n\Delta t).
	\label{eq: cn_def}
\end{equation}
By choosing $n\Delta t = t'$ and $\theta = \omega \Delta t$, in the limit $\Delta t \to 0$ we obtain for the determinant the result
\begin{equation}
	\lim_{t\rightarrow\infty}D(t)=\exp\Bigl [ \frac{t}{2\pi}\int_{-\infty}^{\infty}\!\!\mathrm{d}\omega \ln C(\omega)+\int_0^\infty \!\!\mathrm{d}t' t' g(t')g(-t')\Bigr ],
	\label{eq: det_szego}
\end{equation}
where
\begin{equation}
	g(t') 
	=\frac{1}{2\pi}\int_{-\infty}^{\infty}\mathrm{d}\omega\, \mathrm{e}^{-\mathrm{i}\omega t'}\ln C(\omega)
	\label{eq: g_t_def2}
\end{equation}
and $C(\omega)$ is defined by Eq.~\eqref{eq:Cfw}. Now, the previous condition \eqref{eq: index_def} for the winding number is replaced by a requirement of analyticity of the integrand $\ln C_{\alpha}^f(\omega)$. This means that the real part of $C_{\alpha}^f(\omega)$ must be positive when the imaginary part changes sign; in other case its logarithm will become a multi-valued function, requiring a branch cut. This criterion can be easily visualized by the Nyquist (Cole-Cole) type plots: at half filling and for $U<0.866$, the Nyquist plot does not wind around the origin---that is, the winding number is zero, see Fig.~\ref{fig:fwindingU08}, whereas, for $U>0.866$,  the Nyquist plot does wind around the origin and the winding number is $-1$, see Fig.~\ref{fig:fwindingU10}.

\section{Winding of characteristic function for OTOC}
\label{sec:OTOCwinding}

\begin{figure}[!htb]
	\centering
	\includegraphics[width=0.35\textwidth]{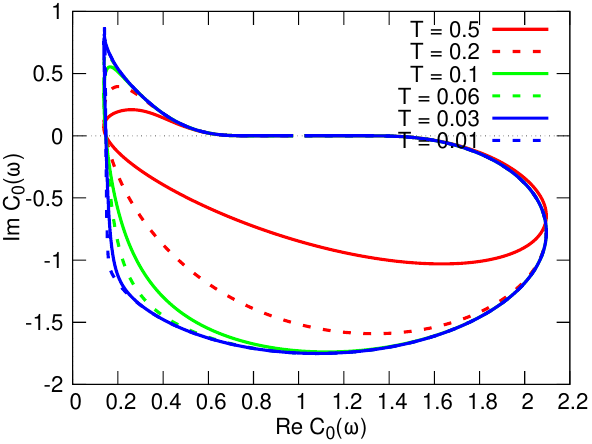}
	\caption{Winding of the characteristic function $C_{0}(\omega)$ for $U=0.4$, $n_f=0.5$, $n_d=0.5$.}
	\label{fig:windingU04}
\end{figure}

\begin{figure}[!tb]
	\centering
	\includegraphics[width=0.35\textwidth]{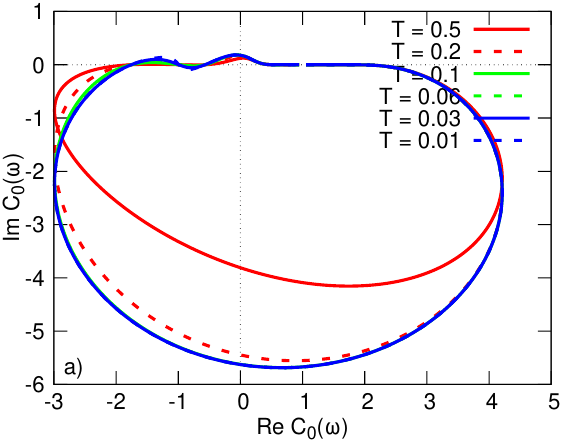} \\
	\includegraphics[width=0.35\textwidth]{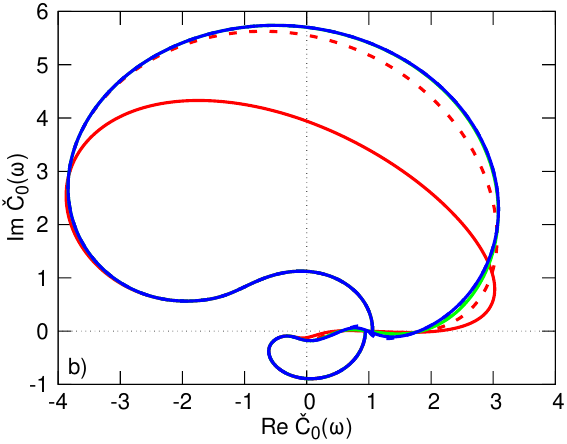}
	\caption{Winding of the characteristic functions (a) $C_{0}(\omega)$ and (b) $\check{C}_{0}(\omega)$ for $U=0.9$, $n_f=0.5$, $n_d=0.5$, and $\Delta t = 0.1$.}
	\label{fig:windingU09}
\end{figure}

\begin{figure}[!tb]
	\centering
	\includegraphics[width=0.35\textwidth]{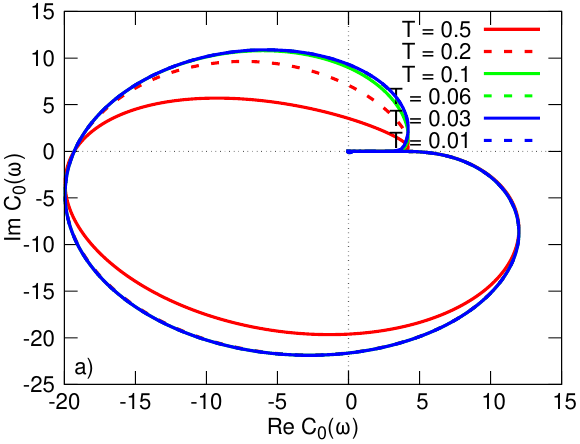} \\
	\includegraphics[width=0.35\textwidth]{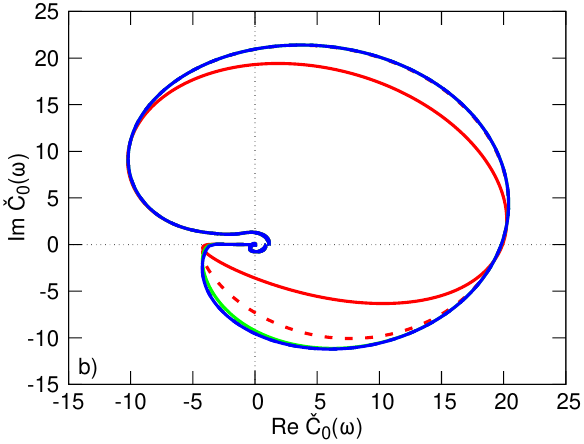}
	\caption{Same as in Fig.~\ref{fig:windingU09} for $U=2.0$.}
	\label{fig:windingU20}
\end{figure}

Similar to the $f$-particle propagator, the long time asymptotic of the determinants in~\eqref{eq:OTOCdet} or $F_{\alpha}(t)$ in \eqref{eq:F01t_fin} is determined according to the Szeg\H{o}-Widom theorem for block Toeplitz matrices~\cite{widom_1976:1} by the same expressions \eqref{eq: szego}--\eqref{eq: g_t_def2}, but with the characteristic functions $C_{0,1}(\omega)$, see Eq.~\eqref{eq:Cw}, 
\begin{align}
	C_{0,1}(\omega) 
	&= \det\nolimits\left[\begin{array}{cc}
		1 \mp U \mathcal{G}_{0,1}^t(\omega) & \mp U \mathcal{G}_{0,1}^{<}(\omega) \\
		\mp U \mathcal{G}_{0,1}^{>}(\omega) & 1 \mp U \mathcal{G}_{0,1}^t(\omega) 
	\end{array}\right]
	\nonumber \\
	&= \left[1 \mp U \operatorname{Re} \mathcal{G}_{0,1}^r(\omega)\right]^2 -U^2 \left[\operatorname{Im}\mathcal{G}_{0,1}^r(\omega)\right]^2
	\nonumber \\
	&\mp 2\mathrm{i} U \tanh\frac{\beta\omega}{2}\operatorname{Im}\mathcal{G}_{0,1}^r(\omega)\left[1 \mp U \operatorname{Re} \mathcal{G}_{0,1}^r(\omega)\right].
    \nonumber \\
	\label{eq:Cw_SW}
\end{align}
Now, the imaginary part of each characteristic function can become zero either at one frequency, $\omega=0$, or at two frequencies, $\omega=0$ and when $1 \mp U \operatorname{Re} \mathcal{G}_{0,1}^r(\omega)=0$. At half filling and for $U<0.44$, the Nyquist plots for $C_{0,1}(\omega)$ do not wind around the origin, Fig.~\ref{fig:windingU04}, and the winding number is zero. For $0.44<U<1.235$, the characteristic functions $C_{0,1}(\omega)$ wind once around the origin as shown in Fig.~\ref{fig:windingU09}, and the winding number is $-1$, whereas for $U>1.235$ it winds twice, as shown in Fig.~\ref{fig:windingU20},---here, the winding number is $-2$.

\section{Winding of the logarithmic derivative}\label{sec:Lyapunov}

\begin{figure}[!tb]
	\centering
	\includegraphics[width=0.3\textwidth]{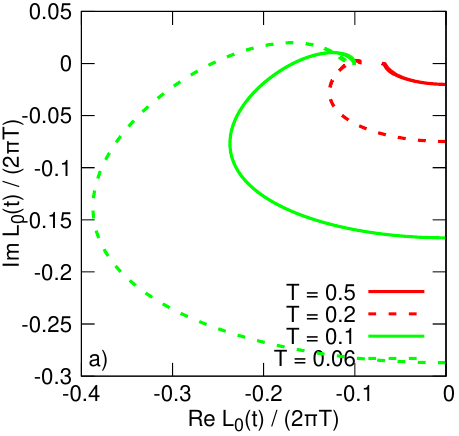} \\ 
	\includegraphics[width=0.3\textwidth]{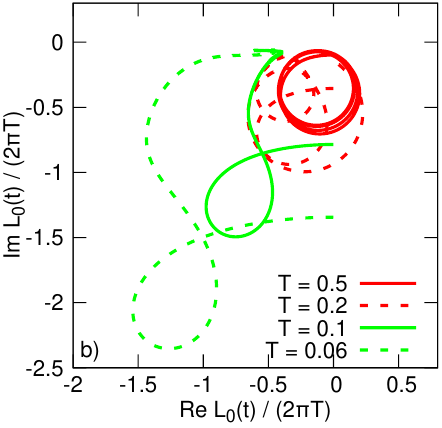} \\
	\includegraphics[width=0.3\textwidth]{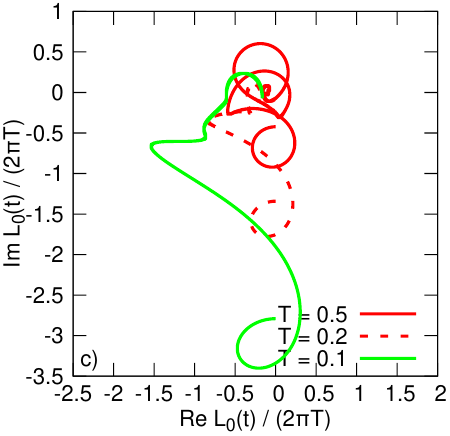} 
	\caption{Winding of the logarithmic derivative $L_0(t)$ for (a) $U=0.4$, (b) $U=0.9$, (c) $U=2.0$, $n_f=0.5$, and $n_d=0.5$.}
	\label{fig:Lyapunov}
\end{figure}

The existence of three different winding numbers for the characteristic function is easily seen on the Nyquist plots for the logarithmic derivative \eqref{eq:Lyapunov} too, see Fig.~\ref{fig:Lyapunov} for an example. For $U<0.44$, the logarithmic derivative smoothly evolves to some steady value. For $0.44<U<0.866$ and at high temperatures, $L_0(t)$ first displays evolution similar to what was seen in the previous case. It then is replaced by circulations, whereas at low temperatures we still observe a smooth evolution to a steady value. For $0.866<U<1.235$, we observe circulations of the logarithmic derivative at high temperatures, which are destroyed (become more chaotic) with decreasing  temperature and are replaced by some complicated looped trajectories. For $U>1.235$, due to the presence of pseudogap and Mott gap, the trajectories become even more chaotic, Fig.~\ref{fig:Lyapunov}.


\providecommand{\noopsort}[1]{}

\end{document}